# Fano Interference in a Single-Molecule Junction

Yiping Ouyang[1,5,#], Rui Wang[1,#], Deping Guo[2,#], Yang-Yang Ju[3,#], Danfeng Pan[4], Xuecou Tu[4], Lin Kang[4], Jian Chen[4], Peiheng Wu[4], Xuefeng Wang[4], Jianguo Wan[1,5], Minhao Zhang[1,5,*], Wei Ji[2,*], Yuan-Zhi Tan[3,*], Su-Yuan Xie[3] and Fengqi Song[1,5,*].

[1]National Laboratory of Solid State Microstructures, Collaborative Innovation Center of Advanced Microstructures, and School of Physics, Nanjing University, Nanjing 210093, China.

[2]Beijing Key Laboratory of Optoelectronic Functional Materials and Micro-Nano Devices, and Department of Physics, Renmin University of China, Beijing 100872, China.

[3]State Key Laboratory of Physical Chemistry of Solid Surfaces, College of Chemistry and Chemical Engineering, Xiamen University, Xiamen 361005, China.

[4]School of Electronic Science and Engineering and Collaborative Innovation Center of Advanced Microstructures, Nanjing University, Nanjing 210023, China.

[5]Atom Manufacturing Institute (AMI), Nanjing 211805, China.

---

[#]These authors contributed equally: Yiping Ouyang, Rui Wang, Deping Guo, Yang-Yang Ju.

[*]Corresponding authors. Email: M.Z.(zhangminhao@nju.edu.cn), W.J.(wji@ruc.edu.cn), Y.T.(yuanzhi_tan@xmu.edu.cn), and F.S.(songfengqi@nju.edu.cn)

# Abstract

Trends of miniaturized devices and quantum interference electronics lead to the long desire of Fano interference in single-molecule junctions, here, which is successfully demonstrated using the 2,7-di(4-pyridyl)-9,9'-spirobifluorene molecule with a long backbone group and a short side group. Experimentally, the two electrically coupled groups are found to contribute to two blurred degenerate points in the differential conductance mapping. This forms a characteristic non-centrosymmetric double-crossing feature, with distinct temperature response for each crossing. Theoretically, we describe the practical in-junction electron transmission using a new two-tunnelling-channel coupling model and obtain a working formula with a Fano term and a Breit-Wigner term. The formula is shown to provide a good fit for all the mapping data and their temperature dependence in three dimensions, identifying the Fano component. Our work thus forms a complete set of evidence of the Fano interference in a single-molecule junction induced by two-tunnelling-channel coupling transport. Density functional theory calculations are used to corroborate this new physics.

**KEYWORD:** Fano interference, quantum interference, single-molecule junctions, a two-tunnelling-channel coupling model, noncentrosymmetrical double-crossing feature, differential conductance mapping

**Introduction**

The question of quantum interference (QI) in single-molecule junctions (SMJs) has attracted considerable attention in terms of the theoretical values of the interference physics at the smallest scale, and its practical utility in microelectronics.[1-5] The control of QI has been demonstrated in various SMJs through the design of molecular structures[6-21], contact configurations[10-12], or electrical gating[12-18]. The molecular structure determines the paths of electron transport, forming the basis of QI in an SMJ. The electrode contact configuration influences the phase difference among electron paths, which is crucial for the effect of QI. The electrical field introduced by a gate can be used to tune the energy levels of molecular states, bringing with it the possibility of changing the status of QI.

Fano interference is a specific and essential type of QI that occurs between a discrete state and a continuous state[22], and is remarkable for the characteristic asymmetrical resonant profile and for the drastic destructive-constructive switching. Having been realized in many mesoscopic electronics devices[23-32], it is now highly sought after in molecular electronics due to its characteristic features[2,5,17-21,32-40]. However, systematic evidence of Fano interference in SMJs is still rare due to both experimental and theoretical challenges. Experimentally, the molecule is difficult to be designed, and be placed perfectly in a practical junction with a complicated environment[40-42]. Theoretically, attempts to distinguish the Fano interference from other types of QIs in SMJs have been made by some pioneers[38,39], but random molecular locations and contact configurations induce a complicated in-junction environment with unexpected interference in some paths, making it extremely hard to describe and identify the characteristics of Fano interference.

We tackle the problems in both experimental and theoretical aspects. We use a 2,7-di(4-pyridyl)-9,9'-spirobifluorene (DPSBF) molecule consisting of a long backbone group and a short side group, where the nitrogen anchoring sites help control the in-junction configuration. We constructed a two-tunnelling-channel coupling model to describe the electronic transmission in practical. We fabricated the field effect transistor devices based on the DPSBF molecule, obtain a characteristic noncentrosymmetrical double-crossing feature (NCDCF) in the differential conductance mapping, and fit the data with their temperature dependence in three dimensions, thus giving a complete set of evidence for the onset of the Fano interference in the SMJ. From density functional theory (DFT), we confirm information about interfered continuous-discrete states from the backbone and the side groups, corroborating the two-tunnelling-channel coupling physics.

**Results and Discussions**

**Fano interference induced by a two-tunnelling-channel coupling model**

The identification of Fano interference relies on a practical model. In a proposed ideal model[2,17-21,34-36,38] (**Figure** 1(a)), the side group of the molecule is coupled only to the backbone group and does not contribute to a tunnelling channel, leading to Fano interference between the side quasi-bond state and the backbone tunnelling state. However, in a real SMJ where the nanogap is rather small, the tunnelling channels through other groups should also exist even in the simplest structure. Therefore, we propose a more practical two-tunnelling-channel coupling model as shown in Fig. 1(b). In this case, the tunnelling channels through the backbone and the side groups coexist, and the side-backbone coupling $u$ leads to the interference between the two channels. The side group has much weaker couplings to the electrodes ($t'_L$ and $t'_R$) than the backbone group ($t_L$ and $t_R$), leading to a discrete tunnelling state to join the Fano interference with a much more continuous backbone tunnelling state.

We describe the electron transmission as a linear superposition of an asymmetrical Fano resonant term, a symmetrical BW resonant term, and an off-resonant term:

$$\tau(E) = A_F \frac{\left(\frac{2(E-\epsilon_F)}{\Gamma_F}+q\right)^2}{\left(\frac{2(E-\epsilon_F)}{\Gamma_F}\right)^2+1} + A_{BW} \frac{1}{\left(\frac{2(E-\epsilon_{BW})}{\Gamma_{BW}}\right)^2+1} + A_{off} \qquad (1)$$

where $q$ is the asymmetry factor of the Fano resonance; $A_F$, $A_{BW}$, and $A_{off}$ are the amplitudes of the three terms; the Fano and BW resonant levels are respectively equal to the effective energy level of the side and backbone groups, i.e., $\epsilon_F = \epsilon_p$ and $\epsilon_{BW} = \epsilon_b$; the Fano and BW resonant widths are equal to the full width at half maximum (FWHM) of the density of states (DOS) of the side and backbone groups, i.e., $\Gamma_F = \Gamma_p$, $\Gamma_{BW} = \Gamma_b$. This form is temperature-independent (when $k_BT \ll \Gamma_p, \Gamma_b$, where $k_B$ is the Boltzmann constant and $T$ is the temperature). Details are given in Supplementary Information S1.

This model considers more factors than the ideal one, describing the inevitable side channel and the interference between channels, but its transmission form is born to distinguish the Fano interference from other types of QI. The Fano and BW resonant terms contribute to a noncentrosymmetrical and a centrosymmetrical crossing patterns in the differential conductance results respectively. This noncentrosymmetrical double-crossing feature (NCDCF) is the characteristic fingerprint for identifying the two-tunnelling-channel coupling induced signal. Moreover, the temperature response of the differential conductance should follow a square law (when $k_B T \ll \varGamma_F, \varGamma_{BW}$). Details are found in Supplementary Information S2.

**A NCDCF and its temperature dependence provide direct evidence of two-tunnelling-channel coupling induced Fano interference**

We designed and synthesized the DPSBF molecule consisting of a long 2,7-di(pyrldin-4-yl)-9H-fluoren backbone and a short 9H-fluoren side group to fabricate the field effect transistor devices (**Figure**. 2(a)). As shown in Fig. 2(b), the nitrogen atoms ensure a robust connection between the backbone group, rather than the side group, and the gold electrodes. This configuration is in accordance with the model. The variation of the junction conductance ($G_{sd}$) is recorded during a feedback-controlled electromigration break junction (FCEBJ) process[43,44] as shown in Fig. 2(c). Further details of the molecular synthesis and device fabrication are provided in the Methods section. The obvious modulation of the Coulomb blockade region in the source-drain current ($I_{sd}$) by the gate voltage ($V_g$) indicates the effectiveness of the gating on the molecular energy levels (Fig. 2(d)). Moreover, the differential conductance ($dI_{sd} / dV_{sd}$) shows several peaks as a function of $V_g$ at the bias voltage $V_{sd} = 0$ mV (Fig. 2(e)). Each peak is related to a degenerate point, namely the condition when the level of a resonant component in the transmission spectrum aligns with the chemical potential of the source and drain electrodes.

In **Figure** 3, we show the characteristic NCDCF as key evidence of the two-tunnelling-channel coupling induced Fano interference. Fig. 3(a) shows the mapping of $dI_{sd}/dV_{sd}$ against $V_g$ and $V_{sd}$ which exhibits an anomalous global noncentrosymmetrical pattern and two adjacent blurred degenerate points at 3 K. The two adjacent degenerate points may result from two coupled tunnelling channels, and the noncentrosymmetry may reflect Fano interference. These features thus provide preliminary evidence of two-tunnelling-channel coupling induced Fano interference.

According to our model, $\epsilon_F$ and $\epsilon_{BW}$ are related to the energy levels of the molecular groups and can be tuned by $V_g$. The parameters $q$, $\Gamma_F$, $\Gamma_{BW}$, $A_F$, $A_{BW}$, and $A_{off}$ are energy-independent and hence they can be obtained via fitting. Hence we apply our model to fit the data (see Eq. (S24) in Supplementary Information) near the degenerate points. As shown in Fig. 3(b), the fitted $dI_{sd}/dV_{sd}$ mapping agrees well with the experimental values, and the signal is attributed to the superposition of a centrosymmetrical BW crossing and a noncentrosymmetrical Fano crossing. This is precisely the characteristic NCDCF as the fingerprint for identifying the two-tunnelling-channel coupling induced signal (Details of individual crossing are provided in Supplementary Information S2, **Figure** S2 and S3). We applied black (white) auxiliary lines to denote the conditions when $\epsilon_F$ ($\epsilon_{BW}$) aligns with the source and drain chemical potentials i.e., the exact locations of the double-crossing in Figs. 3a and b. The fitted parameters are $\alpha = 0.0103$ (where $e\alpha V_g$ is the change in the electrochemical potential caused by the gate electrode and $e$ is the electron charge), $q = 0.374$, $\Gamma_{BW} = 11.91$ meV, $\Gamma_F = 6.79$, $\epsilon_F - \epsilon_{BW} = 8.04$ meV. The asymmetry factor $q = 0.374$ implies that the Fano term behaves more like a dip, rather than a peak of the BW term.

To provide a clearer illustration of this NCDCF, we plot $dI_{sd} / dV_{sd}$ - $V_g$ curves for different $V_{sd}$ under a vertical offset. Figs. 3(c–f) show the experimental curves, the fitting curves with all the components, the BW component from the fitting, and the Fano component from the fitting at different values of $V_{sd}$. The experimental and total fitted data show similar profiles for all $V_{sd}$ values. The BW curves show mirror symmetry, for example, the BW curve at $V_{sd} = +10$ mV is mirror symmetrical to the curve at $V_{sd} = -10$ mV in Fig. 3(e) (plotted in the same colour). In contrast, the Fano curves in Fig. 3(f) show no symmetry and demonstrate the profile of a clear dip with an inconspicuous peak at $V_{sd} = 0$ mV. Thus, we show the distinct difference between the symmetries of the Fano and BW components.

Our model also indicates that the signals of the Fano and BW components will become more unobvious as the temperature increases, reflecting the blunting and widening of peak and dip due to the change of Fermi distribution function while transmission spectrum remains unchanged. There is a linear relation between the $dI_{sd} / dV_{sd}$ values and square of the temperature ($T^2$). When increasing the temperature, the variation tendency of the $dI_{sd} / dV_{sd}$ values is positive near the Fano crossing (as for the dip-like Fano term when $q = 0.374$), and is negative near the BW crossing. All these laws are in the condition that $T \ll \Gamma_F / k_B \approx 50$ K. Further details of temperature response are provided in Supplementary Information S2, **Figures** S4 and S5. In **Figure** 4, we show the temperature-dependent behaviours by constructing the measurements at three different temperatures: 3, 9, and 12K those are much smaller than 50 K. By subtracting the experimental and fitted $dI_{sd}/dV_{sd}$ data against $V_g$ and $V_{sd}$ at 3 K from that at 12 K as plotted in Figs. 4(a) and (b). The black (white) auxiliary lines show that the Fano (BW) component brings a positive (negative) response, in good agreement with the theoretical model.

To obtain more specific details of the temperature response, we compare the experimental $dI_{sd}/dV_{sd}$ - $V_g$ curves for zero bias at 3, 9, and 12 K under a vertical offset in Fig. 4c. As the temperature increases, there are different responses in the BW (peak) and the Fano (dip) regions. In these two regions, for illustrative purposes we defined two obvious points on the curve as point J at $V_g$ = 2.85 V and point X at $V_g$ = 2.00 V. Point J is near the peak position of the curve, while X is near the dip position on the right side of J. $X_{Temp.}$ and $J_{Temp.}$ are the values of $dI_{sd}/dV_{sd}$ expressed in units of $e^2/h$ unit at 3, 9, and 12 K. The temperature dependence of $X_{Temp.}$ and $J_{Temp.}$ shows an opposite trend (Fig. 4(g)), and presents a linear relationship with $T^2$. Figs. 4d-f show fitted results with total and individual contributions of the BW and Fano terms at three different temperatures. The tendencies of $X_{Temp.}$ and $J_{Temp.}$ in Fig. 4(h) collected from the fitted data in Fig. 4(d) agree well with the experimental results in Fig .4(g), while Figs. 4(i) and (j) show the two distinct tendencies corresponding to the contribution of the BW peak and the Fano dip, respectively.

As described above, our data show the characteristic NCDCF in the $dI_{sd}/dV_{sd}$ mapping, with each crossing showing a distinct temperature responses but both following a square law. All of these agree well with our model in three dimensions, forming the complete set of evidence of the two-tunnelling-channel coupling induced Fano interference. This level of agreement demonstrates that the two-tunnelling-channel coupling model is effective in describing the practical in-junction environment, and distinguish the Fano interference. This model has been extended to studies on Fano interference in device 2, provided in Supplementary Information S3, and **Figure** S7.

**Density functional theory calculations corroborate the two-tunnelling-channel coupling model**

Our model pays particular attention to the spatial configurations of the SMJ. The side and backbone channels should respectively lead to discrete and continuous states in the Fano interference. Information on the positions and widths of the states are reflected in the results of $dI_{sd} / dV_{sd}$. The position of related state levels should be distributed in an appropriate range to be accessed by gating. All these features are confirmed by our DFT calculations based on the structure illustrated in **Figure** 5a. Due to the high differential conductance that we measured ($\sim 10^{-1}$ $e^2/h$ near the maximum $dI_{sd} / dV_{sd}$), we consider the backbone group to be highly overlapped with electrodes and to lie across the nanogap, while the side group hangs between it. Further details of the calculations are provided in Methods section.

The DOS profile of DFT calculations (Fig. 5d) corresponding to the fitted zero-bias $dI_{sd} / dV_{sd}$ - $\alpha eV_g$ curves (Fig. 5e) is observed around the Fermi level ($E_f$). The reason why the energy range is not exactly at the Fermi level may be the difference of gold electrodes between the theoretical model and the experimental environment, such as the surface relaxation and the adsorption of adatom. Both the DOS of the backbone and the side groups have a central peak separated in energy and are labelled as $P_1$ and $P_2$ respectively. The partial charge densities at $P_1$ and $P_2$ are shown in Figs. 5(b) and (c), presenting distinctly different distributions of the wave functions in the SMJ. The FWHM of $P_1$ and $P_2$ are $\Gamma_b$ = 12.07 meV and $\Gamma_p$ = 6.41 meV, while the corresponding fitted results give $\Gamma_{BW}$ = 11.91 meV $\approx \Gamma_b$ and $\Gamma_F$ = 6.79 meV $\approx \Gamma_p$. Furthermore, the distance between $P_1$ and $P_2$ is $\epsilon_p - \epsilon_b$ = 8.70 meV, which is in good agreement with the distance between the BW and Fano resonant levels in the fitted results $\epsilon_F - \epsilon_{BW}$ = 8.04 meV. Note that $\epsilon_p$, $\epsilon_b$, $\Gamma_p$, and $\Gamma_b$ all relate to the states after considering the side-backbone interaction but not simply to the original discrete and continuous tunnelling states, i.e. $\Gamma_p$ is not much smaller than $\Gamma_b$ because the final side state is broadened not only by $t'_L$ and $t'_R$ but also by $u$. Thus the DFT results give a corroboration of the physics of two-tunnelling-channel coupling induced Fano interference

**Conclusions**

In summary, we have obtained the systematic evidence of Fano interference in an SMJ via both theoretical and experimental approaches. Theoretically, we proposed a practical two-tunnelling-channel coupling model to describe and identify the signal of Fano interference in SMJs. Experimentally, we fabricated field effect transistor devices based on the DPSBF molecule to obtain the characteristic NCDCF in the $dI_{sd} / dV_{sd}$ mapping, noting that the temperature dependence follows a square law. The DFT calculations also corroborate the physics. All these results demonstrate the existence of two-tunnelling-channel coupling induced Fano interference in an SMJ. Our model and methods of identification could be extended into further studies of Fano interference in SMJs, and promote the realization of practical QI in single-molecule electronics.

**Methods**

**Molecule preparation and characterization**

2,7-dibromo-9,9'-spiro-bifluorene (474 mg, 1.0 mmol), Pd(dppf) Cl$_2$(67 mg, 0.10 mmol), 4-pyridinylboronic acid (368 mg, 3.0 mmol) and K$_2$CO$_3$ (690 mg, 5.0 mmol) in the mixture of tetrahydrofuran (10 mL) and water (2 mL) was stirred in Argon atmosphere at 75 ℃ for 36 h. After cooling it to room temperature, the reaction mixture was concentrated under a reduced pressure and purified using silica gel column chromatography with petroleum ether/ethyl acetate (*v/v*, 3:1) as the eluent to afford DPSBF (249 mg, 53%) as a white solid. We performed the nuclear magnetic resonance (NMR) measurements for molecular characterization, and all details are shown in Supplementary information S3.

**Device Preparation**

The gold nanowires (about 50-nm-width) were deposited via electron beam evaporation on top of a silicon gate electrode that has previously been added a SiO$_2$ gate dielectric layer (about 30-nm-thick) via atomic layer deposition. The external circuits and SiO$_2$ layer were patterned using an ultra-violet lithography procedure, but the nanowire was defined by electron-beam lithography. The array were cleaned further using oxygen plasma after the gold nanowires were prepared. Then the device containing DPSBF molecules was cooled to 1.6 K, and a nanogap was produced using the FCEBJ method[43,44]. The current was monitored while the source-drain voltage was increased. The voltage was decreased to 10 mV as soon as the current dropped by 1%. The cycle is repeated until the conductance was < 0.025 $e^2/h$ for the voltage of 20 mV.

**Density functional theory calculations**

Our DFT calculations were performed using the generalized gradient approximation and the projector augmented wave method[45,46] as implemented in the Vienna ab-initio simulation package[47]. The kinetic energy cut-off was set to 400 eV. The Γ point was used to sample the first Brillouin zone in all calculations. All atoms of the molecule were allowed to relax until the residual force on each atom was less than 0.01 eV Å$^{-1}$. The Grimme's D3 form van der Waals (vdW) correction was considered with the Perdew–Burke–Ernzerhof (PBE) exchange functional (PBE-D3).[48,49] The Dual-Slab Model[50] was adopted, consisting of five layers of Au atoms for each slab and separated by a vacuum region. The molecule on the Au (111) interface was located between two slabs with mirror symmetry. All the vacuum layer adopted (> 10 Å) is sufficient to appreciably reduce the image interactions.

**Conflicts of interest**

The authors declare no conflicts of interests.

**Acknowledgements**

We are most grateful to Professor Mark Reed for inspiring this work, and for many years of cooperation, and we acknowledge the financial support of the National Key R&D Program of China (No. 2017YFA0303203 and No. 2018YFE0202700), the National Natural Science Foundation of China (Grant Nos. 92161201, 12104221,

12104220, 12025404, 12004174, 11904165, 11904166, 61822403, 11874203 and 11974422), the Strategic Priority Research Program of Chinese Academy of Sciences (Grant No. XDB30000000), the Natural Science Foundation of Jiangsu Province (Grant Nos. BK20200312, BK20200310, and BK20190286), the Fundamental Research Funds for the Central Universities (Grant No. 020414380192), and the Fundamental Research Funds for the Central Universities, and the Research Funds of Renmin University of China (Grant Nos. 22XNKJ30 and 21XNH090).

**Contributions**

M.Z. and F.S. conceived the research, W.J., and Y.-Z.T. co-supervised the project. D.G. and W.J. performed and analysed the density functional theory calculations. Y.O. fabricated the devices and performed the SMJ measurements. R.W. proposed the theoretical model. Y.-Z.T. participated in the preparation of the molecular materials. Y.-Y.J. prepared the molecular materials. X.T. and D.P. assisted in the device fabrication. Y.O., R.W., D.G., M.Z., F.S., and W.J. wrote the paper. L.K., J.C., P.W., X.W., J.W. and S.-Y.X. participated in discussions on this manuscript. All authors discussed the results and commented on the manuscript.

**Figure Captions**

**Figure 1. Two possible Fano interference models for the nanojunction of a single molecule with both a backbone group and a side group**

**(a)** This ideal model describes the Fano interference between a quasi-bound side state ($\epsilon_p$) and a tunnelling backbone state ($\epsilon_b$) when the side group is coupled only with the backbone group ($u$) in an SMJ. **(b)** This two-tunnelling-channel coupling model describes the more practical Fano interference in an SMJ. Besides the interactions in the ideal model described in **(a)**, the electrons from the left electrode may also tunnel through the side group state to the right because of the random location of the molecule in the nanojunction with a rather small space. This case is rather practical and general in SMJs and leads to the coexistence of backbone and side channels along with their interference. The coupling between the side group and electrodes ($t'_L$, $t'_R$) is much smaller than that between the backbone group and electrodes ($t_L$, $t_R$) hence resulting in much a smaller broadening on the side state than on the backbone state. The Fano resonance occurs between the discrete (narrow) side tunnelling state and the continuous (broad) backbone state. This two-tunnelling-channel coupling model contains a Fano resonant term linear superposition to the BW resonant term in the transmission spectrum as formulated by Eq. (1). The levels ($\epsilon_F$, $\epsilon_{BW}$) and widths ($\Gamma_F$, $\Gamma_{BW}$) of two resonances are respectively determined by the effective levels ($\epsilon_p$, $\epsilon_b$) and FWHMs of side and backbone states ($\Gamma_p$, $\Gamma_b$).

**Figure 2. Basic measurements of the transistor device based on the SMJ of DPSBF**

**(a)** Typical configuration of the transistor device based on the SMJ. Two independent variables $V_g$ and $V_{sd}$ are tunable. **(b)** Local configuration of the molecular attachment, where the molecule DPSBF is bonded onto the electrodes by the N connection. **(c)** $G_{sd}$ - $V_{sd}$ curves between source and drain electrodes during the breaking process to fabricate the junction via a FCEBJ technique. A deeper colour implies longer time. The bias conductance reduced over time until to a target value. **(d)** $I_{sd}$ - $V_{sd}$ curves for different $V_g$ after the FCEBJ process, indicating Coulomb blockade and gate tunability. **(e)** $dI_{sd} / dV_{sd}$ - $V_g$ curve at the zero bias voltage, reflecting the distributions of the DOS.

**Figure 3. The characteristic NCDCF in the mapping of $dI_{sd} / dV_{sd}$**

**(a, b)** Experimental and fitted $dI_{sd} / dV_{sd}$ mapping, in which the characteristic NCDCF can be seen. The black (white) auxiliary lines are shown to respectively mark the positions of $\epsilon_F$ ($\epsilon_{BW}$), where they align with the chemical potentials of source and drain electrodes. **(c–f)** $dI_{sd} / dV_{sd}$ - $V_g$ curves at different $V_{sd}$ as marked on the right, adapted from the data from panels **(a)** and **(b)**. **(c)** experimental curves, **(d)** fitting curves with all the components, **(e)** BW component from the fitting, and **(f)** Fano interference component from the fitting. All the curves are vertically off-set. In **(e)** the mirror symmetry can be seen, characteristic of the BW tunnelling process, e.g., the data are symmetrical at $V_{sd} = +10$ and $-10$ mV. In **(f)**, there is a dip-like profile for $V_{sd} = 0$ mV and total asymmetry for higher bias voltages, which is typical for Fano interference. All data were collected at 3 K.

**Figure 4. Temperature-dependent behaviours of the NCDCF confirm the physics of the two-tunnelling-channel coupling induced Fano interference**

**(a, b)** The difference obtained by subtracting the $dI_{sd} / dV_{sd}$ mapping data at 3 K from those at 12 K: **(a)** experimental data, **(b)** fitting results. The red colour indicates that the values at 3 K are larger than that at 12 K, while the blue colour means they are smaller. The black (white) auxiliary lines show the conditions when $\epsilon_F$ ($\epsilon_{BW}$) aligns with the chemical potential of the source and drain electrodes. **(c–f)** $dI_{sd}/dV_{sd}$ - $V_g$ curves at zero bias voltage at different temperatures as marked on the right, **(c)** experimental data, **(d)** fitting result, **(e)** BW components obtained from the fitting, **(f)** Fano interference components obtained from the fitting, which are vertically off-set. $X_{Temp.}$ ($J_{Temp.}$) shown in the panels relates to the values of $dI_{sd}/dV_{sd}$ in units of $e^2/h$ unit of the Fano (BW) components at the given temperatures. The panels **(g, h, i, j)** show the data of X and J extracted from the panels **(c, d, e, f)** respectively, which are plotted against the square of the temperatures. The temperature responses follow a square law.

**Figure 5. Practical model corroborated by DFT calculations**

**(a)** The proposed configuration of the molecule in the junction, in which the molecular backbone group lies across a nanogap on a gold nanowire and the side group hangs between the nanogap. Note that a rather large part of the backbone group is in strong contact with the gold electrodes. **(b)** and **(c)** show the partial charge density corresponding to the DOS peak contributed from the backbone ($P_1$) and side ($P_2$) group respectively, as marked in **(d)**. **(d)** The computed DOS of the backbone and side groups are shown and the energy is relative the Fermi level ($E_f$). The data of backbone and side groups have different central peaks, which are labelled as P1 and P2 respectively. The position distance and FWHMs of the peaks are also shown. **(e)** shows the fitted $dI_{sd}/dV_{sd}$ - $\alpha eV_g$ curves at zero bias voltage and 3 K extracted from Fig. 3. The corresponding parameters in **(d)** and **(e)** exhibit a good agreement, i.e. $\epsilon_F - \epsilon_{BW} \approx \epsilon_p - \epsilon_b$, $\Gamma_F \approx \Gamma_p$, $\Gamma_{BW} \approx \Gamma_b$, corroborating the physics of the two-tunnelling-channel coupling model.

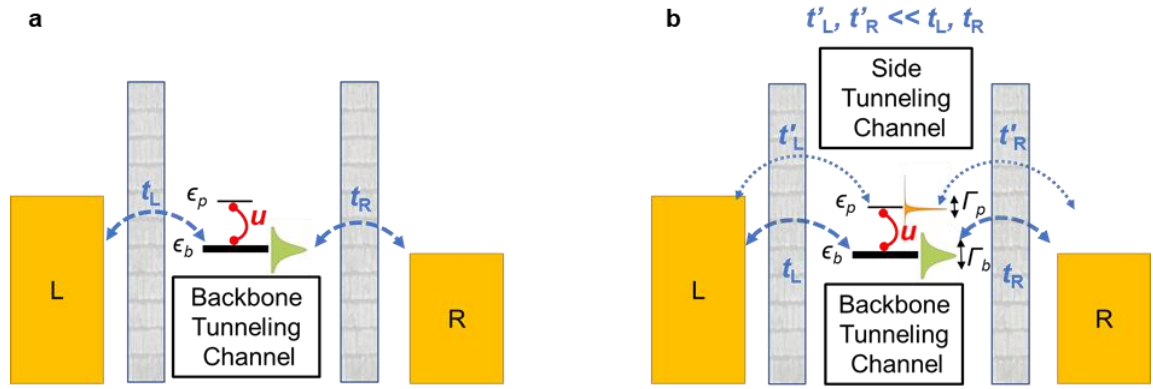

**Ouyang et al. Figure 1**

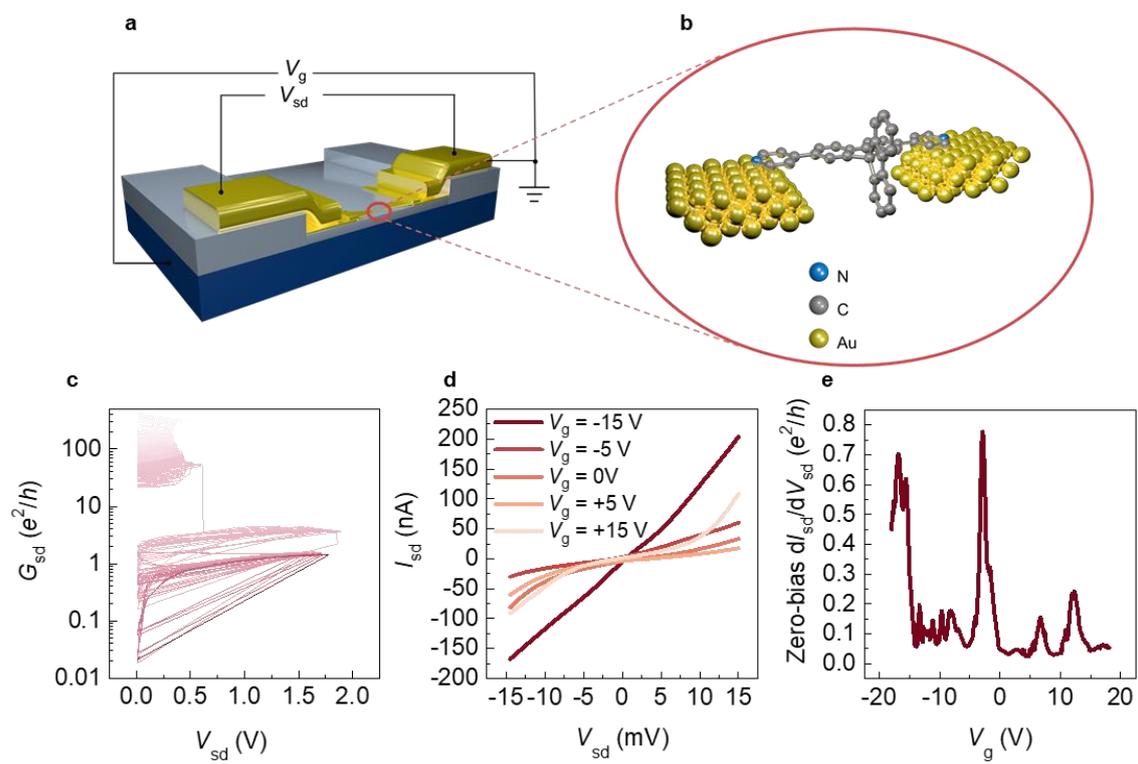

**Ouyang et al. Figure 2**

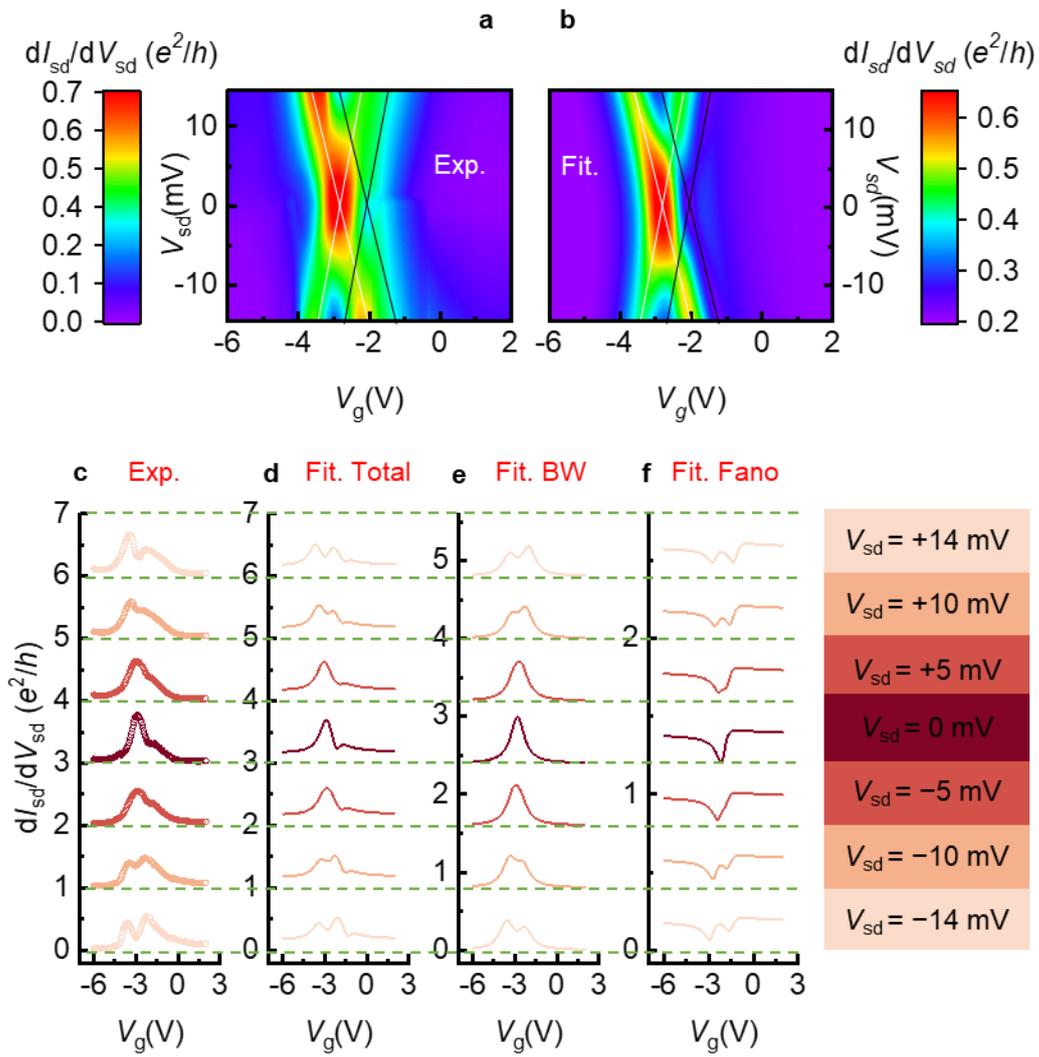

Ouyang et al. Figure 3

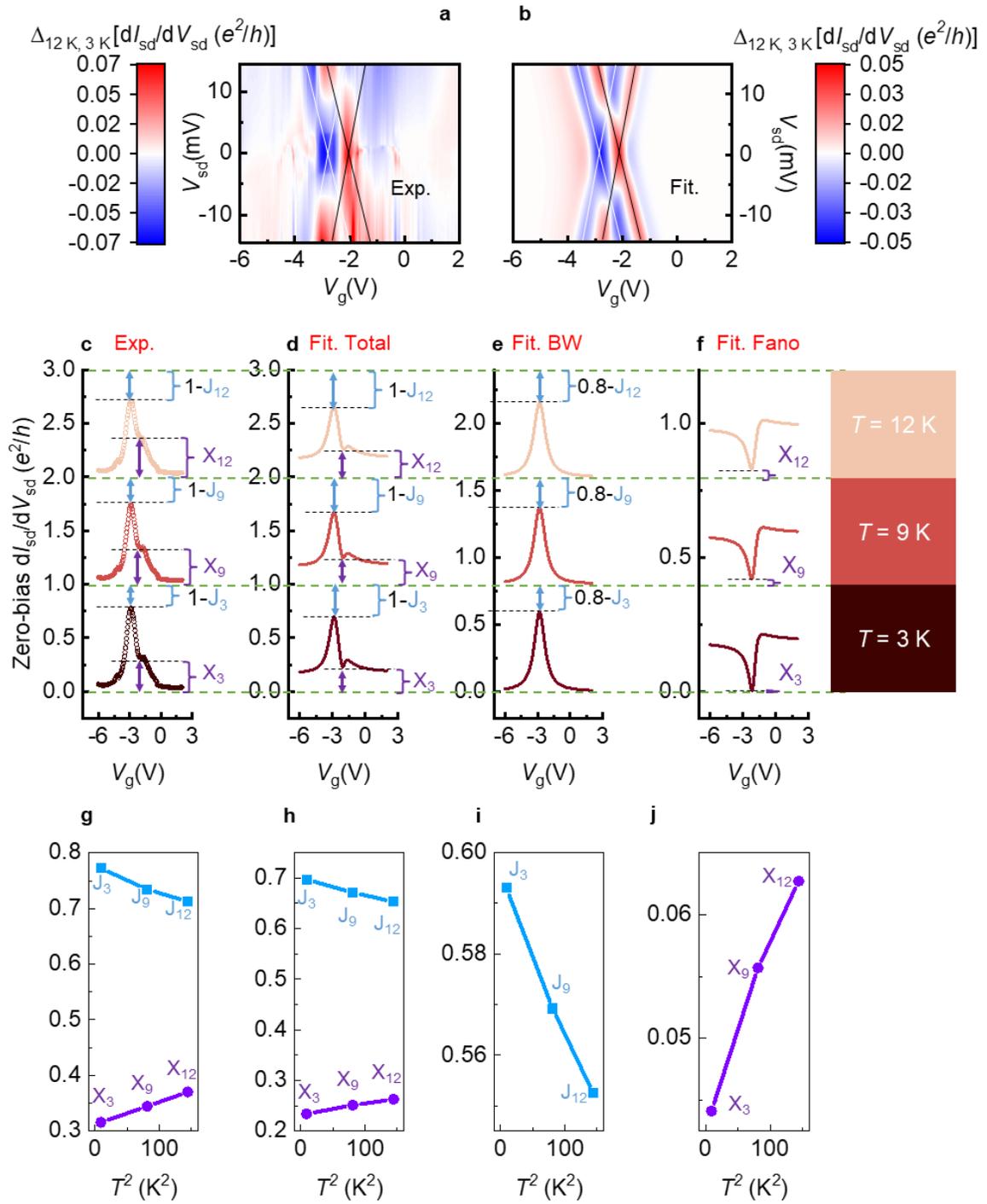

Ouyang et al. Figure 4

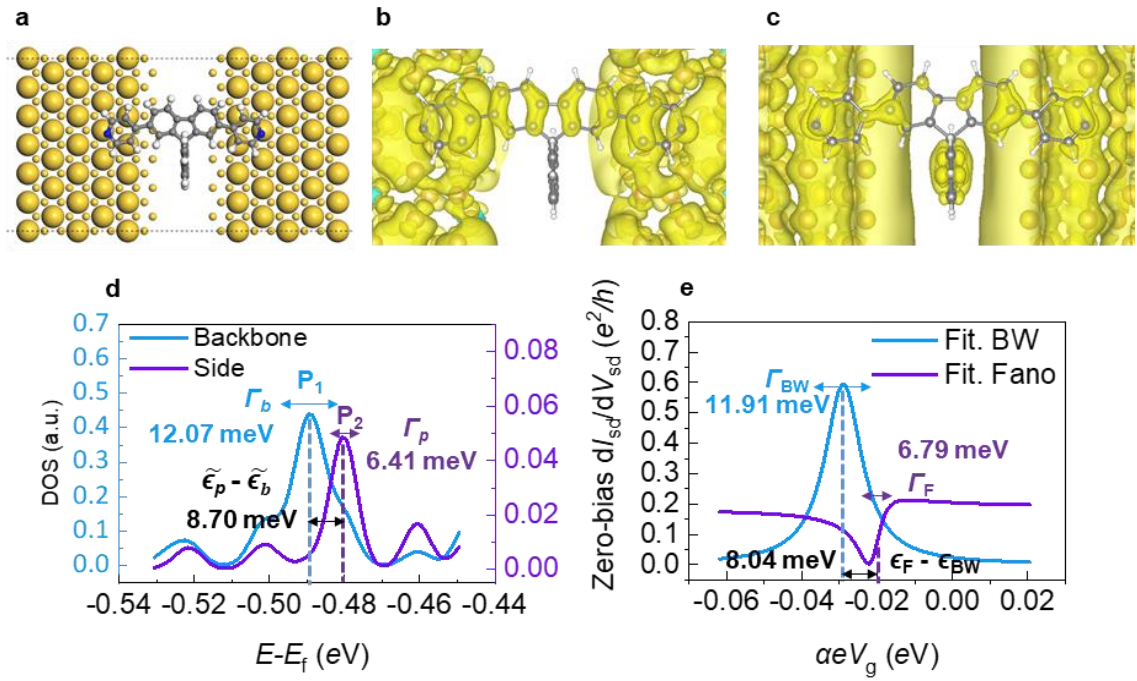

**Ouyang et al. Figure 5**

# Table of Contents

## 1. The Two-tunneling-channel Coupling Model of a Single-molecule Junction

### 1.1 Model and setup

We depict the single-molecule junction (SMJ) with several subparts as shown in Fig. S1: the left electrode $L$, the right electrode $R$, the backbone group $b$, and the side group $p$. Here $b$ is directly coupled to the electrodes with tunneling coefficients $t_L, t_R$. While $p$ has indirect tunneling coefficients $t_L', t_R'$, describing the second-order hopping process. It is natural to assume $t' \ll t$. The intramolecular coupling between $b$ and $p$ is described as a direct coupling constant $u$.

The Hamiltonian of the entire molecular junction is given by:

$$H = H_0 + H_c \tag{S1}$$

Where $H_0$ describes the molecule itself, and $H_c$ describes its couplings to $L$ and $R$.

$$H_0 = \varepsilon_b b^\dagger b + \varepsilon_p p^\dagger p + u \sum (b^\dagger p + h.c.) \tag{S2}$$

$$\begin{aligned} H_c = &\sum_k \left(\varepsilon_{L,k} R_k^\dagger R_k + \varepsilon_{R,k} L_k^\dagger L_k\right) \\ &+ \sum_k t_L(L_k^\dagger b + h.c.) + \sum_k t_R(R_k^\dagger b + h.c.) \\ &+ \sum_k t'_L(L_k^\dagger p + h.c.) + \sum_k t'_R(R_k^\dagger p + h.c.) \end{aligned} \tag{S3}$$

### 1.2 The current through the single-molecule junction

The current through the junction is obtained from $I_{LR} = e\langle \dot{N}_L \rangle = -e\langle \dot{N}_R \rangle = I_{Lm} = I_{mR}$. The mark $m$ stands for the entire molecule. Utilizing the Heisenberg equation of motion, one arrives at

$$I_{Lm} = \frac{2\pi e i}{h}\left[t_L\sum_k(\langle L_k^\dagger b\rangle - \langle b^\dagger L_k\rangle) + t'_L\sum_k(\langle L_k^\dagger p\rangle - \langle p^\dagger L_k\rangle)\right] \quad (S4)$$

Where the spin degrees of freedom are implicit. Making expansion of the two correlation functions to the leading order of $t, t'$, we obtain

$$\begin{aligned}I_{Lm} = \frac{2e}{h}2\pi\rho_L\int_{-\infty}^{\infty}dE\,&[f(E-\mu_L)-f(E-\mu_m)]\\ \times\,&[t_L'^2(G_p^a-G_p^r)+t_Lt_L'(G_{pb}^a-G_{pb}^r)\\ +\,&t_L't_L(G_{bp}^a-G_{bp}^r)+t_L^2(G_b^a-G_b^r)]\end{aligned} \quad (S5)$$

Here $f(\omega) = \left(\exp\left(\frac{\omega}{k_BT}\right)+1\right)^{-1}$ is the Fermi distribution function, $k_B$ is the Boltzmann constant, $\mu_L, \mu_m$ are the chemical potentials of the left electrode and the molecule, $\rho_L$ is the density of states (DOS) of the left electrode, $G_b^{r,a}$ and $G_p^{r,a}$ are retarded or advanced Green's functions (GFs) of $b$ and $p$, $G_{pb}^{r,a}$ and $G_{bp}^{r,a}$ are the crossed GF. The GFs are expressed as

$$G_b^{r,a}(E) = \frac{1}{E-\varepsilon_b-\Sigma_b^{r,a}(E)} \quad (S6)$$

$$G_p^{r,a}(E) = \frac{1}{E-\varepsilon_p-\Sigma_p^{r,a}(E)} \quad (S7)$$

$$G_{pb}^{r,a} = G_{bp}^{r,a} = ug_b^{r,a}G_p^{r,a} = ug_p^{r,a}G_b^{r,a} \quad (S8)$$

Where $\Sigma_b^{r,a}(E), \Sigma_p^{r,a}(E)$ are the self-energy of $b$ and $p$, describing perturbations from other components, read as:

$$\Sigma_b^{r,a}(E) = g_L^{r,a}t_L^2 + g_R^{r,a}t_R^2 + g_p^{r,a}u^2 \quad (S9)$$

$$\Sigma_p^{r,a}(E) = g_L^{r,a}t_L'^2 + g_R^{r,a}t_R'^2 + g_b^{r,a}u^2 \quad (S10)$$

$g_b^{r,a} = \frac{1}{E-\varepsilon_b-(g_L^{r,a}t_L^2+g_R^{r,a}t_R^2)} = \frac{1}{E-\varepsilon_b-\sigma_b^{r,a}}$ and $g_p^{r,a} = \frac{1}{E-\varepsilon_p-(g_L^{r,a}t_L'^2+g_R^{r,a}t_R'^2)} = \frac{1}{E-\varepsilon_p-\sigma_p^{r,a}}$ are the GFs of $b$ and $p$ without $b$-$p$ interaction but already consider the broadening from the electrodes, with the self-energy $\sigma_b^{r,a} = g_L^{r,a}t_L^2 + g_R^{r,a}t_R^2$ and $\sigma_p^{r,a} = g_L^{r,a}t_L'^2 +$

$g_R^{r,a} t_R'^2$; $g_L^{r,a}$ and $g_R^{r,a}$ are GFs of electrodes that are nearly unaffected by the couplings from the molecule, for $\text{Im}[g_L^{r,a}] = \mp i\pi\rho_L, \text{Im}[g_R^{r,a}] = \mp i\pi\rho_R$, $\text{Re}[g_L^{r,a}] \propto \mu_L$, $\text{Re}[g_R^{r,a}] \propto \mu_R$. Here we have modeled both $b$ and $p$ by an effective single quantum state.

By using Dyson's equation, GF of backbone could be written as

$$G_b^a = g_b^a + g_b^a u G_{pb}^a = g_b^{r,a} + u^2 g_b^{a^2} G_p^a \tag{S11}$$

Hence Eq. (S5) leads to

$$\begin{aligned} I_{Lm} &= \frac{2e}{h} 4\pi\rho_L \int_{-\infty}^{\infty} dE \, [f(E-\mu_L) - f(E-\mu_m)] \\ &\quad \times \text{Im}[t_L^2(g_b^a) + (g_b^{a^2} t_L^2 u^2 + 2t_L t_L' u g_b^a + t_L'^2) G_p^a] \\ &= \frac{2e}{h} 4\pi\rho_s \int_{-\infty}^{\infty} dE \, [f(E-\mu_L) - f(E-\mu_m)] \\ &\quad \times \text{Im}[t_L^2 g_b^a + (t_L u g_b^a + t_L')^2 G_p^a] \\ &= \frac{2e}{h} 4\pi\rho_s \int_{-\infty}^{\infty} dE \, [f(E-\mu_L) - f(E-\mu_m)] \times \zeta(E) \end{aligned} \tag{S12}$$

Where $\zeta(E) = \text{Im}[t_L^2 g_b^a + (t_L u g_b^a + t_L')^2 G_p^a]$.

**1.3 The origin of the Fano term in the transmission**

We consider the couplings to the left and right electrodes are approximately equal, i.e. $t_L' \approx t_R' = t'$, $t_L \approx t_L = t$; and the DOS of the left and right electrodes are the same constant, i.e. $\rho_L \approx \rho_R = \rho$. And then by introducing

$$q = \frac{t' + tu\text{Re}[g_b^a]}{tu\text{Im}[g_b^a]}, \epsilon_p = \varepsilon_p + \text{Re}[\Sigma_p^a] \tag{S13}$$

$\zeta(E)$ could be written as

$$\zeta(E) = \text{Im}\left[t_L^2 g_b^a + \frac{(tu\text{Im}[g_b^a])^2}{\text{Im}[\Sigma_p^a]}\left(\frac{\left(\frac{E-\epsilon_p}{\text{Im}[\Sigma_p^a]} + q\right)^2}{\left(\frac{E-\epsilon_p}{\text{Im}[\Sigma_p^a]}\right)^2 + 1} - 1\right)\right] \tag{S14}$$

On account of that $t' \ll t$, the broadening from electrodes of $p$ is neglected, i.e. $\text{Im}[\Sigma_p^a] = t'^2 \text{Im}[g_L^a] + t'^2 \text{Im}[g_R^a] + u^2 \text{Im}[g_b^a] \approx u^2 \text{Im}[g_b^a]$. We also consider $t$ is large enough so the $b$ state is well broadened to be a quasi-continuum. Therefore the term $\dfrac{(tu\text{Im}[g_b^a])^2}{\text{Im}[\Sigma_p^a]} \left( \dfrac{\left(\frac{E-\widetilde{\epsilon_p}}{\text{Im}[\Sigma_p^a]}+q\right)^2}{\left(\frac{E-\widetilde{\epsilon_p}}{\text{Im}[\Sigma_p^a]}\right)^2+1} - 1 \right)$ describes the interference between a discrete $p$ state and a quasi-continuum $b$, i.e. in this term the real and imaginary parts of backbone GF are approximately constant, for $\text{Im}[g_b^a] \approx \text{Im}\left[\dfrac{1}{\sigma_b^a}\right] = \dfrac{1}{2\rho\pi t^2}$, $\text{Re}[g_b^a] \approx \text{Re}\left[\dfrac{1}{\sigma_b^a}\right]$. Note that $g_b^a$ in the individual term $t^2 g_b^a$ is still expressed as $\dfrac{1}{E-\epsilon_b-\sigma_b^a}$. We further introduce $\epsilon_F = \epsilon_p$, $\Gamma_F = 2\text{Im}[\Sigma_p^a] \approx 2u^2\text{Im}[g_b^a]$, $\epsilon_{BW} = \epsilon_b + \text{Re}[\sigma_b^{r,a}] = \epsilon_b$, $\Gamma_{BW} = 2\text{Im}[\sigma_b^{r,a}] = 4\rho\pi t^2$, now parameters $\epsilon_F$, $\Gamma_F$, $\epsilon_{BW}$, $\Gamma_{BW}$, and $q$ are all independent to energy, and Eq. (S14) is written as:

$$\zeta(E) = t^2\left(-\frac{\Gamma_F}{2u^2} + \frac{2}{\Gamma_{BW}}\frac{1}{\left(\frac{E-\epsilon_{BW}}{\Gamma_{BW}}\right)^2+1} + \frac{\Gamma_F}{2u^2}\frac{\left(\frac{E-\epsilon_F}{\Gamma_F}+q\right)^2}{\left(\frac{E-\epsilon_F}{\Gamma_F}\right)^2+1}\right) \quad (S15)$$

To adopt Landauer's formula[1]

$$I_{LR} = \frac{2e}{h}\int_{-\infty}^{\infty} dE\, \tau(E)[f(E-\mu_L) - f(E-\mu_R)] \quad (S16)$$

The transmission is given by

$$\tau(E) = 2\pi t^2 \rho \left\{ -\frac{\Gamma_F}{2u^2} + \frac{2}{\Gamma_{BW}}\frac{1}{\left(\frac{2(E-\epsilon_F)}{\Gamma_{BW}}\right)^2+1} + \frac{\Gamma_F}{2u^2}\frac{\left(\frac{2(E-\epsilon_F)}{\Gamma_F}+q\right)^2}{\left(\frac{2(E-\epsilon_F)}{\Gamma_F}\right)^2+1} \right\} \quad (S17)$$

It could be written as the form mentioned in the manuscript:

$$\tau(E) = A_F \frac{\left(\frac{2(E-\epsilon_F)}{\Gamma_F}+q\right)^2}{\left(\frac{2(E-\epsilon_F)}{\Gamma_F}\right)^2+1} + A_{BW}\frac{1}{\left(\frac{2(E-\epsilon_F)}{\Gamma_{BW}}\right)^2+1} + A_{\text{off}} \quad (S18)$$

Where the amplitudes of three terms are:

$$A_{\text{off}} = 2\pi t^2 \rho \left(-\frac{\Gamma_{\text{F}}}{2u^2}\right), A_{\text{BW}} = 2\pi t^2 \rho \frac{2}{\Gamma_{\text{BW}}}, A_{\text{F}} = 2\pi t^2 \rho \frac{\Gamma_{\text{F}}}{2u^2} \qquad (S19)$$

Now we have revealed the physics of this two-tunneling-channel coupling model containing a Fano term, a BW term and an off-resonance term in the transmission. In general, the side state has much narrower broadening from the electrodes than that of backbone state. Therefore, the role of discrete state and continuous state in the Fano interference are played by the side and backbone state respectively. The interference is described in the Fano term, while the BW term is roughly only related to the backbone tunneling alone. All the parameters in the transmission are independent with temperature when $k_B T \ll \Gamma_p, \Gamma_b$, because the electrical coupling factors are irrelevantly to temperature and the broadening from thermal energy is neglected. The Fano and BW resonant levels $\epsilon_{\text{F}} = \epsilon_p$ and $\epsilon_{\text{BW}} = \epsilon_b$ are the effective energy levels of side and backbone states; while the resonant widths $\Gamma_{\text{F}} = \Gamma_p$ and $\Gamma_{\text{BW}} = \Gamma_b$ are the full widths at half maximum (FWHM) of DOS of side and backbone states after considering the b-p interaction. $\Gamma_p$ would be much larger than the FWHM of DOS of original discrete side tunneling state because the broadening from coupling $u$ has be taken into account.

We design the 2,7-di(4-pyridyl)-9,9'-spirobifluorene (DPSBF) molecule to achieve the configuration in this model: firstly, it consists of a long 2,7-di(pyrldin-4-yl)-9H-fluoren backbone group, connected with a 9H-fluoren side group via a single carbon atom, it is reasonable to regard the backbone and side groups as two semi-independent subparts; secondly, the backbone group is longer than side one, and possessing nitrogen anchor on each end, ensuring a mighty enough $t$ while keeping the side group an appropriate distance from the electrodes so that $t'$ is small.

## 2 Fitting details by the model

### 2.1 Tunability of the gate electrode in the device

In the field-effect transistor devices based on the SMJ, there are three electrodes, the source ($s$), the drain ($s$). and the gate ($g$). We define the source as the left electrode, and the grounded drain as the right electrode, hence

$$\mu_L = \mu_s = -eV_{sd} \qquad \mu_R = \mu_d = 0 \qquad (S20)$$

We consider the whole molecule have common capacitances between three electrodes, therefore the molecular levels are shifted by voltages with the same factor:

$$\epsilon_F = \epsilon_p = \epsilon_{p0} - \eta eV_{sd} - \alpha eV_g, \epsilon_{BW} = \epsilon_b = \epsilon_{b0} - \eta eV_{sd} - \alpha eV_g \qquad (S21)$$

Where $\epsilon_{p0}$ and $\epsilon_{b0}$ are the levels when $V_{sd} = 0$, $V_g = 0$. The factors $\eta = \frac{C_s}{C_s + C_d + C_g}$ and $\alpha = \frac{C_g}{C_s + C_d + C_g}$ are related to capacitances between the molecule and electrodes.

### 2.2 The double-crossing feature on differential conductance mapping

We then derive the characteristic noncentrosymmetrical double-crossing feature (NCDCF) originating from the superposition of Fano and BW components in our two-tunneling-channel coupling model.

The current given by Landauer's formula is written as

$$I_{sd} = \frac{2e}{h} \int_{-\infty}^{\infty} dE \, [A_F \frac{\left(\frac{E - (\epsilon_{p0} - \eta eV_{sd} - \alpha eV_g)}{\Gamma_F} + q\right)^2}{\left(\frac{E - (\epsilon_{p0} - \eta eV_{sd} - \alpha eV_g)}{\Gamma_F}\right)^2 + 1}$$
$$+ A_{BW} \frac{1}{\left(\frac{E - (\epsilon_{b0} - \eta eV_{sd} - \alpha eV_g)}{\Gamma_{BW}}\right)^2 + 1} \qquad (S22)$$
$$+ A_{off}] \times [f(E + eV_{sd}) - f(E)]$$

If the temperature is 0 K, differential conductance $\frac{dI_{sd}}{dV_{sd}}$ has an analytical solution:

$$\left.\frac{dI_{sd}}{dV_{sd}}\right|_{T=0} = \frac{2e^2}{h}[A_{\text{off}}$$

$$+ A_{\text{BW}}\left(\frac{\eta}{\left(\left(\frac{\epsilon_{\text{BW}}}{\Gamma_{\text{BW}}}\right)^2 + 1\right)} + \frac{1-\eta}{\left(\left(\frac{eV_{sd} - \epsilon_{\text{BW}}}{\Gamma_{\text{BW}}}\right)^2 + 1\right)}\right)$$

$$+ A_{\text{F}}\left(1 - \frac{(1-q^2)\eta}{\left(\left(\frac{\epsilon_{\text{F}}}{\Gamma_{\text{F}}}\right)^2 + 1\right)} - \frac{2q\eta\frac{\epsilon_{\text{F}}}{\Gamma_{\text{F}}}}{\left(\frac{\epsilon_{\text{F}}}{\Gamma_{\text{F}}}\right)^2 + 1}\right.$$

$$\left. - \frac{(1-q^2)(1-\eta)}{\left(\left(\frac{eV_{sd} + \epsilon_{\text{F}}}{\Gamma_{\text{F}}}\right)^2 + 1\right)} - \frac{2q(1-\eta)\frac{\epsilon_{\text{F}}}{\Gamma_{\text{F}}}}{\left(\frac{eV_{sd} + \epsilon_{\text{F}}}{\Gamma_{\text{F}}}\right)^2 + 1}\right)] \quad (S23)$$

Similar to the form of the transmission, it is the superposition of a Fano term, a BW term and an off-resonant term. For zero bias condition,

$$\left.\frac{dI_{sd}}{dV_{sd}}\right|_{T=0,V_{sd}=0} = \frac{2e^2}{h}[A_{\text{off}} + A_{\text{BW}}\left(\frac{1}{\left(\left(\frac{\epsilon_{b0} - \alpha eV_g}{\Gamma_{\text{BW}}}\right)^2 + 1\right)}\right)$$

$$+ A_{\text{F}}\left(\frac{\left(\frac{\epsilon_{p0} - \alpha eV_g}{\Gamma_{\text{F}}} - q\right)^2}{\left(\left(\frac{\epsilon_{p0} - \alpha eV_g}{\Gamma_{\text{F}}}\right)^2 + 1\right)}\right)] \quad (S24)$$

Compare Eq. (S24) with Eq. (S18), it can be seen that $\left.\frac{dI_{sd}}{dV_{sd}}\right|_{T=0,V_{sd}=0}$ as the function of $\alpha eV_g$ has the same line shape with $\tau(E)$, Therefore changing the gate voltage is a direct method to extract the feature of Fano interference in the transmission.

Equation (S23) gives $\frac{dI_{sd}}{dV_{sd}}$ as the function of $V_g$ and $V_{sd}$, hence we could analyze the theoretical results on $V_g$-$V_{sd}$ mapping, for total signal and individual Fano (BW) component.

As for the BW term, the $\frac{dI_{sd}}{dV_{sd}}$ pattern on the mapping is always a centrosymmetrical crossing, the factor $\eta$ only influence the slope of the crossing edges, examples of mapping with two different $\eta$ values are plotted in Fig. S2(a) and (b), along with the corresponding $\frac{dI_{sd}}{dV_{sd}}$ curves against $V_g$ at different $V_{sd}$ are plotted in Fig. S2(c) and (d). While diverse noncentrosymmetrical crossings of $\frac{dI_{sd}}{dV_{sd}}$ could be produced by Fano term as shown in Fig. S3, due to the codetermination from $q$ and $\eta$.( Fig. S3(a-d) for mappings and (e-h) for $\frac{dI_{sd}}{dV_{sd}}$ curves against $V_g$)

In our model, $\frac{dI_{sd}}{dV_{sd}}$ is the sum of three terms and only the Fano term contributes to the noncentrosymmetrical ingredient (the off-resonant term only produces an invariable background). Therefore, the total signal exhibit a pattern with characteristic NCDCF.

Our experimental $\frac{dI_{sd}}{dV_{sd}}$ result (see Fig. 3(a)) show an obvious noncentrosymmetry with two blurred degenerate points, therefore we fit the data with the two-tunneling-channel coupling model to see whether it could be expressed as the NCDCF.

**2.3 Fit the data by the approximation at 0 K**

Indeed, we construct our experiments at a nonzero temperature of 3 K, this low enough temperature only would cause a tiny deviation from the zero-Kelvin results, does not influence the identification by the NCDCF. Therefore, we directly fit the data at 3 K with Eq. (S24) via the Curve Fitting tool in MATLAB 2021a. We obtain all the fitted parameters:

$$A_{\text{off}} = -0.0057, A_{\text{BW}} = 0.2997, A_{\text{F}} = 0.0934, \eta = 0.5846, \alpha = 0.0103$$

$$\Gamma_{\text{BW}} = 11.91 \text{ meV}, \Gamma_{\text{F}} = 6.79 \text{ meV}, \epsilon_{\text{F}} - \epsilon_{\text{BW}} = 8.04 \text{ meV}, q = 0.374$$

The analysis of fitted results are described in the manuscript text and Figs. 3, in which the components of Fano and BW are both identified.

**2.4 The temperature dependence of the differential conductance**

The fitted results in 3 K give $\Gamma_{BW} = 11.91$ meV, $\Gamma_F = 6.79$ meV, and consider $\Gamma_F = \Gamma_p$ and $\Gamma_{BW} = \Gamma_b$, hence condition of the temperature-independence of the transmission: $k_B T \ll \Gamma_p, \Gamma_b$ is satisfied. Therefore the theoretical $\frac{dI_{sd}}{dV_{sd}}$ at higher temperatures is obtained by numerical integration of Eq. (S22), in which the parameters are the same as that obtained in 3 K fitting.

The linear relation between $\frac{dI_{sd}}{dV_{sd}}$ variation and $T^2$ is derived from the Sommerfeld expansion when $k_B T \to 0$:

$$\int_{-\infty}^{\infty} \frac{H(E)}{\exp\left(\frac{(E-\mu)}{k_B T}\right) + 1} dE$$

$$= \int_{-\infty}^{\mu} H(E) dE + \frac{\pi^2}{6} (k_B T)^2 \left.\frac{dH(E)}{dE}\right|_{E=\mu} \quad \text{(S25)}$$

$$+ O\left(\frac{k_B T1}{\mu}\right)^4$$

Neglect the higher orders and substitute $\tau(E)$ in Eq. (S18) into $H(E)$, the current in a finite temperature $T$ is obtained as:

$$I_{LR} = \frac{2e}{h}\left[\int_{-\infty}^{\mu_L} \tau(E) dE + \frac{\pi^2}{6}(kT)^2 \left.\frac{d\tau(E)}{dE}\right|_{E=\mu_L}\right]$$

$$- \frac{2e}{h}\left[\int_{-\infty}^{\mu_R} \tau(E) dE + \frac{\pi^2}{6}(kT)^2 \left.\frac{d\tau(E)}{dE}\right|_{E=\mu_R}\right] \quad \text{(S26)}$$

$$= \frac{2e}{h}\int_{-\infty}^{\infty} dE\, \tau(E)[f_0(E-\mu_L) - f_0(E-\mu_R)]$$

$$+ \frac{2e}{h}\frac{\pi^2}{6}(kT)^2 \left(\left.\frac{d\tau(E)}{dE}\right|_{E=\mu_L} - \left.\frac{d\tau(E)}{dE}\right|_{E=\mu_R}\right)$$

Where $f_0(\omega) = \lim\limits_{T \to 0}\left(\exp\left(\frac{\omega}{k_B T}\right) + 1\right)^{-1}$ is the Fermi distribution function in 0 K, therefore the deviation of $\frac{dI_{sd}}{dV_{sd}}$ from that in 0 K is $\frac{2e}{h}\frac{\pi^2}{6}(kT)^2 \frac{d}{dV_{sd}}\left(\frac{d\tau(E)}{dE}\bigg|_{E=\mu_L} - \frac{d\tau(E)}{dE}\bigg|_{E=\mu_R}\right)$. This deviation is proportional to $T^2$ when $T < \frac{\Gamma_F}{k_B} \approx 50$ K because $\tau(E)$ is temperature-independent. The temperature response reflects the blunting and widening of peak and dip due to the change of Fermi distribution function while transmission spectrum remains unchanged. This square law are only in the condition that $T \ll \Gamma_F / k_B \approx 50$ K, the signal would not maintain increasing or decreasing if $T$ is too higher.

To investigate the temperature-dependent behaviors, we measure $\frac{dI_{sd}}{dV_{sd}}$ at three different temperatures. For comparison, we have translated the $V_g$ coordinates of $\frac{dI_{sd}}{dV_{sd}}$ obtained from different times of measurements to align the peak positions of zero-bias $\frac{dI_{sd}}{dV_{sd}}$. In the manuscript text, the experimental and fitted $\frac{dI_{sd}}{dV_{sd}}$ difference between at 3 K and 12 K along with the variation of zero-bias $\frac{dI_{sd}}{dV_{sd}}$ as the temperature increases are shown, here we discuss more details about the temperature dependence of individual Fano and BW components.

The temperature response of $\frac{dI_{sd}}{dV_{sd}}$ for individual BW and Fano term are plotted in Fig. S4 and Fig S5. As shown in Fig. S4a and b, the difference of $\frac{dI_{sd}}{dV_{sd}}$ between at 12 K and 3 K of the BW component is always centrosymmetrical and exhibits as blue crossing (means negative temperature response), because of the symmetrical BW peak is simply become gentler when temperature increase, the details are described in Fig. S4e and f via the same method in the manuscript by setting the point J$_{Temp}$ at $V_g = -2.85$ V, note that different $\eta$ lead to same curve at zero-bias. The Fano result shows some diverse response in different $q$ and $\eta$ values. For our fitted result $q = 0.374$, the Fano term is roughly a dip. As shown in Fig. S5 a and b, difference of $\frac{dI_{sd}}{dV_{sd}}$ between 12 K and 3 K exhibits as nearly a red crossing (means positive temperature response), but for some other result things may be more complex, e.g., when $q = 1.374$, individual Fano component would bring two overlapped crossings pattern due to a typical dip-peak transmission profile (see Fig. S5 c and d). The variation of curves in different temperature is illustrated in Fig. S5 e-I, here for $q = 0.374$ the point near $X_{Temp}$ the dip is still at $V_g = -2.00$ V, while for $q = 1.374$, both the point near the dip $X_{Temp}$ at $V_g = -2.60$ V and near the peak $Y_{Temp}$ at $V_g = -1.70$ V.

## 3 Additional experimental details

### 3.1 Molecular synthesis and characterization

Synthesis route of DPSBF molecule is shown in Fig. S6a. Nuclear magnetic resonance (NMR) spectra were acquired on a Bruker AV 500 Spectrometer (Germany) at 298 K in the solvents indicated. Chemical shifts are expressed in ppm units relative to TMS (0.00 ppm, 1H). Silica gel (300-400 mesh) was used for column chromatography. All chemicals and solvents were purchased from commercial sources were used without further purification.

NMR spectra of 2,7-di(4-pyridyl)-9,9'-spirobifluorene molecule is shown in Figs. S6b and c, with the following results: $^1$H NMR (500 MHz, CDCl$_3$) δ 8.53 (d, 4H), 7.99 (d, 2H), 7.90 (d, 2H), 7.70 (m, 2H), 7.41 (t, 2H), 7.34 (d, 4H), 7.14 (t, 2H), 7.00 (d, 2H), 6.80(d, 2H) ppm. $^{13}$C NMR (125 MHz, CDCl$_3$) δ 192.69, 150.56, 146.79, 144.28, 139.62, 135.39, 133.51, 123.07, 121.42, 121.20 ppm.

### 3.2 Experimental evidences of Fano interference in device 2

We fabricate some more SMT devices based on 2,7-di(4-pyridyl)-9,9'-spirobifluorene molecules, and obtained another experimental data that agree with our model.

Figure S7 shows the experimental and fitted result of Device 2 at 2 K. In Fig. S7a, the experimental data show two crossing-like $\frac{dI_{sd}}{dV_{sd}}$ patterns on $V_g - V_{sd}$ map, for the left one is roughly centrosymmetrical but the right one is obviously noncentrosymmetrical. By the fitting of our model (Fig, S7b), we consider that the left and right pattern is mainly induced by the BW term and the Fano term respectively. In Figs. S7a and b, the black (white) auxiliary lines are applied to denote the conditions when $\epsilon_F$ ($\epsilon_{BW}$) align with the source and drain electrodes.

The fitted parameters are: $A_{off} = -0.0002$, $A_{BW} = 0.1305$, $A_F = 0.0423$, $\eta = 0.52$, $\alpha = 0.0202$, $\Gamma_{BW} = 7.0$ meV, $\Gamma_F = 5.2$ meV, $q = -1.70$, $\epsilon_F - \epsilon_{BW} = 28.3$ meV. The factor $q = -1.70$ indicates an obvious dip-peak pair in the Fano profile.

Figures S7c-f show experimental curves, the fitting curves with all the components, the BW resonance component from the fitting and the Fano interference component. The symmetry of BW and the asymmetry of Fano component are clearly illustrated.

**Figure Captions**

**Figure S1. Setup of the two-tunneling-channel coupling model**

This model is to describe the practical electron transmission of SMJ based on a molecule possessing both a backbone group and a side group.

The SMJ consists of: the left electrode $L$, the right electrode $R$, the backbone group $b$, and the side group $p$. The couplings between $p$ and electrodes $t'_L$ and $t'_R$ are much weaker than couplings between $b$ and electrodes $t_L$ and $t_R$. The coupling between $b$ and $p$ is $u$.

There are two tunneling channels respectively through the backbone and the side group. The interference between the quasi continuous backbone tunneling state and the relatively discrete side state leads to the Fano resonance in the transmission spectrum.

**Figure S2. Theoretical centrosymmetrical crossing feature of BW component in the differential conductance**

(a) The theoretical $\frac{dI_{sd}}{dV_{sd}}$ against $V_g$ and $V_{sd}$ of BW term based on fitted parameters mentioned in the manuscript: $\eta = 0.5846, \alpha = 0.0103, \Gamma_{BW} = 11.91$ meV.

(b) The theoretical $\frac{dI_{sd}}{dV_{sd}}$ against $V_g$ and $V_{sd}$ of BW term based on parameters: $\eta = 0.4154, \alpha = 0.0103, \Gamma_{BW} = 11.91$ meV.

The white auxiliary lines mark the positions of $\epsilon_{BW}$, where it aligns with the chemical potentials of source and drain electrodes.

The patterns exhibit as crossing that are centrosymmetrical about the intersection point of auxiliary lines, different $\eta$ only lead to different slope of crossing edges.

(c, d) $\frac{dI_{sd}}{dV_{sd}}$ - $V_g$ curves at different $V_{sd}$ as marked on the right, adapted from the data of panels (a) and (b). In (c) and (d) one might see the mirror symmetry, characteristic for the BW tunneling process, e.g. the data are nearly symmetrical at $V_{sd} = +10$ and $-10$ mV.

**Figure S3. Theoretical noncentrosymmetrical crossing feature of Fano component in the differential conductance**

(a) The theoretical $\frac{dI_{sd}}{dV_{sd}}$ against $V_g$ and $V_{sd}$ of Fano term based on fitted parameters mentioned in the manuscript: $\eta = 0.5846, \alpha = 0.0103, \Gamma_F = 6.79$ meV, $q = 0.374$.

(b) The theoretical $\frac{dI_{sd}}{dV_{sd}}$ against $V_g$ and $V_{sd}$ of Fano term based on parameters: $\eta = 0.4154, \alpha = 0.0103, \Gamma_F = 6.79$ meV, $q = 0.374$.

(c) The theoretical $\frac{dI_{sd}}{dV_{sd}}$ against $V_g$ and $V_{sd}$ of Fano term based on n parameters: $\eta = 0.5846, \alpha = 0.0103, \Gamma_F = 6.79$ meV, $q = 1.374$.

(d) The theoretical $\frac{dI_{sd}}{dV_{sd}}$ against $V_g$ and $V_{sd}$ of Fano term based on parameters: $\eta = 0.4154, \alpha = 0.0103, \Gamma_F = 6.79$ meV, $q = 1.374$.

The black auxiliary lines mark the positions of $\epsilon_F$, where it aligns with the chemical potentials of source and drain electrodes. The patterns are all noncentrosymmetrical but with different specific distribution forms, codetermined by $q$ and $\eta$.

(e–h) $\frac{dI_{sd}}{dV_{sd}}$ - $V_g$ curves at different $V_{sd}$ as marked on the right, adapted from the data of panels (a)—(d), all curves are vertically off-set. In (e)—(h), it is a dip at $V_{sd} = 0$ mV and totally noncentrosymmetrical at higher voltages, typically for the Fano interference.

**Figure S4. Theoretical temperature response of BW component in the differential conductance**

(a) The difference of the theoretical $\frac{dI_{sd}}{dV_{sd}}$ difference against $V_g$ and $V_{sd}$ between 12 K and 3 K of BW term, based on fitted parameters mentioned in the manuscript: $\eta = 0.5846, \alpha = 0.0103, \Gamma_{BW} = 11.91$ meV.

(b) The difference of the theoretical $\frac{dI_{sd}}{dV_{sd}}$ difference against $V_g$ and $V_{sd}$ between 12 K and 3 K of BW term, based on fitted parameters: $\eta = 0.4154, \alpha = 0.0103, \Gamma_{BW} = 11.91$ meV. The white auxiliary lines mark the positions of $\epsilon_{BW}$, where it aligns with the chemical potentials of source and drain electrodes.

The patterns manifest as a blue centered crossing near the auxiliary lines, which means the negative temperature response.

(c) $\frac{dI_{sd}}{dV_{sd}}$ - $V_g$ curves at zero bias voltage at different temperatures as marked on the right, which are vertically off-set. J$_{Temp}$ shown in the panels is the values of $dI_{sd} / dV_{sd}$ in $e^2/h$ unit of the BW component at the given temperatures. Different $\eta$ correspond to the same curve on zero bias voltage. The panels (d) is the data of J extracted from the panel (c), which are plotted against the square of the temperatures. All the above data corroborate the physics of two-tunneling-channel coupling induced Fano Interference.

**Figure S5. Theoretical temperature response of Fano term in the differential conductance**

(a) The difference of the theoretical $\frac{dI_{sd}}{dV_{sd}}$ difference against $V_g$ and $V_{sd}$ between 12 K and 3 K of Fano term, based on fitted parameters mentioned in the main body text: $\eta = 0.5846, \alpha = 0.0103, \Gamma_F = 6.79$ meV, $q = 0.374$.

(b) The difference of the theoretical $\frac{dI_{sd}}{dV_{sd}}$ difference against $V_g$ and $V_{sd}$ between 12 K and 3 K of BW term, based on fitted parameters: $\eta = 0.4154, \alpha = 0.0103, \Gamma_F = 6.79$ meV, $q = 0.374$.

(c) The difference of the theoretical $\frac{dI_{sd}}{dV_{sd}}$ difference against $V_g$ and $V_{sd}$ between 12 K and 3 K of BW term, based on fitted parameters: $\eta = 0.5846, \alpha = 0.0103, \Gamma_F = 6.79$ meV, $q = 0.374$.

(d) The difference of the theoretical $\frac{dI_{sd}}{dV_{sd}}$ difference against $V_g$ and $V_{sd}$ between 12 K and 3 K of BW term, based on fitted parameters: $\eta = 0.4154, \alpha = 0.0103, \Gamma_F = 6.79$ meV, $q = 1.374$.

The black auxiliary lines mark the positions of $\epsilon_F$, where it aligns with the chemical potentials of source and drain electrodes.

The patterns show diverse distributions: in (a) and (b), $q = 0.374$ means that the Fano line shape is roughly a dip so the pattern is roughly a red centered crossing near the auxiliary lines, which means the positive temperature response; in (c) and (d), $q = 1.374$ means that the Fano line shape shows both a peak and a dip obviously, so the pattern shows both a red and blue crossing on each side of the auxiliary lines, which means a complex temperature response.

(e) $\frac{dI_{sd}}{dV_{sd}}$ - $V_g$ curves at zero bias voltage at different temperatures as marked on the right, which are vertically off-set. $X_{Temp}$ shown in the panels is the values of $dI_{sd} / dV_{sd}$ in $e^2/h$ unit of the Fano component with $q = 0.374$ at the given temperatures. Different $\eta$ correspond to the same curve on zero bias voltage. The panels **(f)** is the data of X extracted from the panel **(e)**, which are plotted against the square of the temperatures.

(g) $\frac{dI_{sd}}{dV_{sd}}$ - $V_g$ curves at zero bias voltage at different temperatures as marked on the right, which are vertically off-set. $X_{Temp}$ and $Y_{Temp}$ shown in the panels is the values of $dI_{sd} / dV_{sd}$ in $e^2/h$ unit of the Fano component with $q = 1.374$ at the given temperatures. Different $\eta$ correspond to the same curve on zero bias voltage. The panels **(h)** and **(i)** is the data of X and Y extracted from the panel **(g)**, which are plotted against the square of the temperatures.

All the above data corroborate the physics of two-tunneling-channel coupling induced Fano Interference.

### Figure S6, Molecular synthesis and characterization

**(a)** Synthesis route of 2,7-di(4-pyridyl)-9,9'-spirobifluorene molecule.
**(b)** $^1$H NMR spectrum of 2,7-di(4-pyridyl)-9,9'-spirobifluorene molecule, the results are: 8.53 (d, 4H), 7.99 (d, 2H), 7.90 (d, 2H), 7.70 (m, 2H), 7.41 (t, 2H), 7.34 (d, 4H), 7.14 (t, 2H), 7.00 (d, 2H), 6.80(d, 2H) ppm.
**(c)**$^{13}$C NMR spectrum of 2,7-di(4-pyridyl)-9,9'-spirobifluorene molecule, the results are: 192.69, 150.56, 146.79, 144.28, 139.62, 135.39, 133.51, 123.07, 121.42, 121.20 ppm.

### Figure S7, Differential conductance data of device B

**(a)** Experimental $\frac{dI_{sd}}{dV_{sd}}$ results of device B. The left pattern is roughly centrosymmetrical and the right pattern is obviously noncentrosymmetrical.

**(b)** Theoretical $\frac{dI_{sd}}{dV_{sd}}$ result based on the parameters with the fitting by our two-tunneling-channel coupling model. The left pattern mainly originates from the BW term, and the right pattern mainly originates from the Fano term.

The black (white) auxiliary lines are shown to respectively marks the positions of $\epsilon_F$ ($\epsilon_{BW}$), where they align with the chemical potentials of source and drain electrodes.

**(c-f)** $dI_{sd} / dV_{sd}$ - $V_g$ curves at different points of $V_{sd}$ as marked on the right, adapted from the data of panels **(a)** and **(b)**. **(c)** experimental curves, **(d)** the fitting curves with all the components, **(e)** the BW component from the fitting and **(f)** the Fano interference component from the fitting. All the curves are vertically off-set. In **(e)** one might see the mirror symmetry, characteristic for the BW tunneling process, e.g. the data are symmetrical at $V_{sd} = +8$ and $-8$ mV. In **(f)**, it is a dip-peak pair at $V_{sd} = 0$ and totally asymmetrical at higher voltages, typically for the Fano interference. All the data are collected at 2 K.

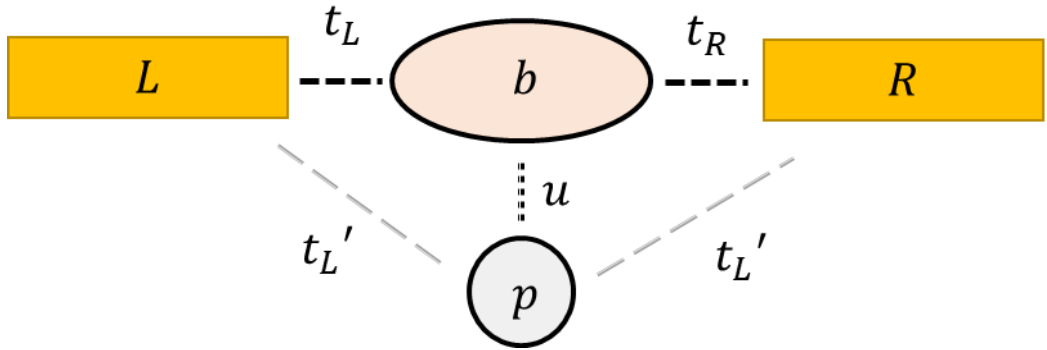

**Ouyang et al. Figure S1**

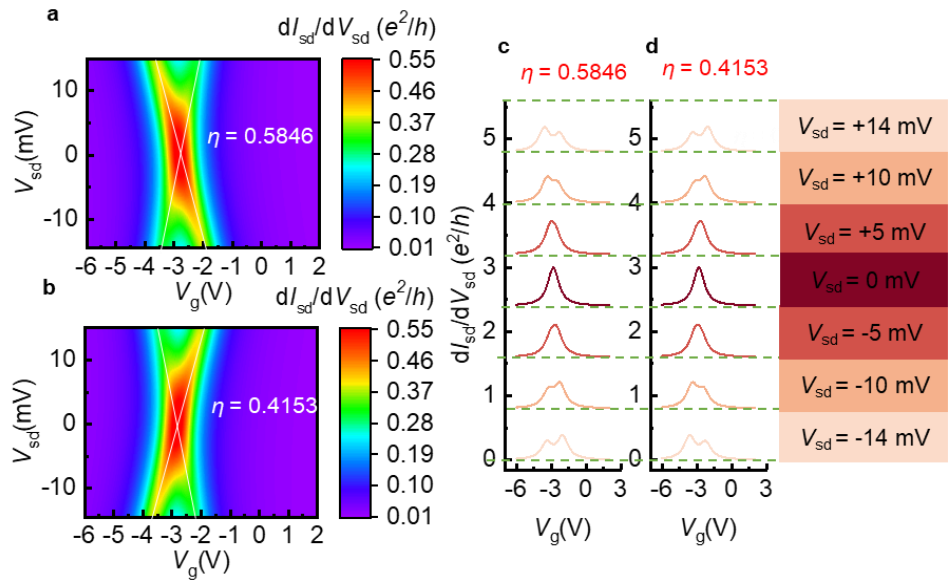

**Ouyang et al. Figure S2**

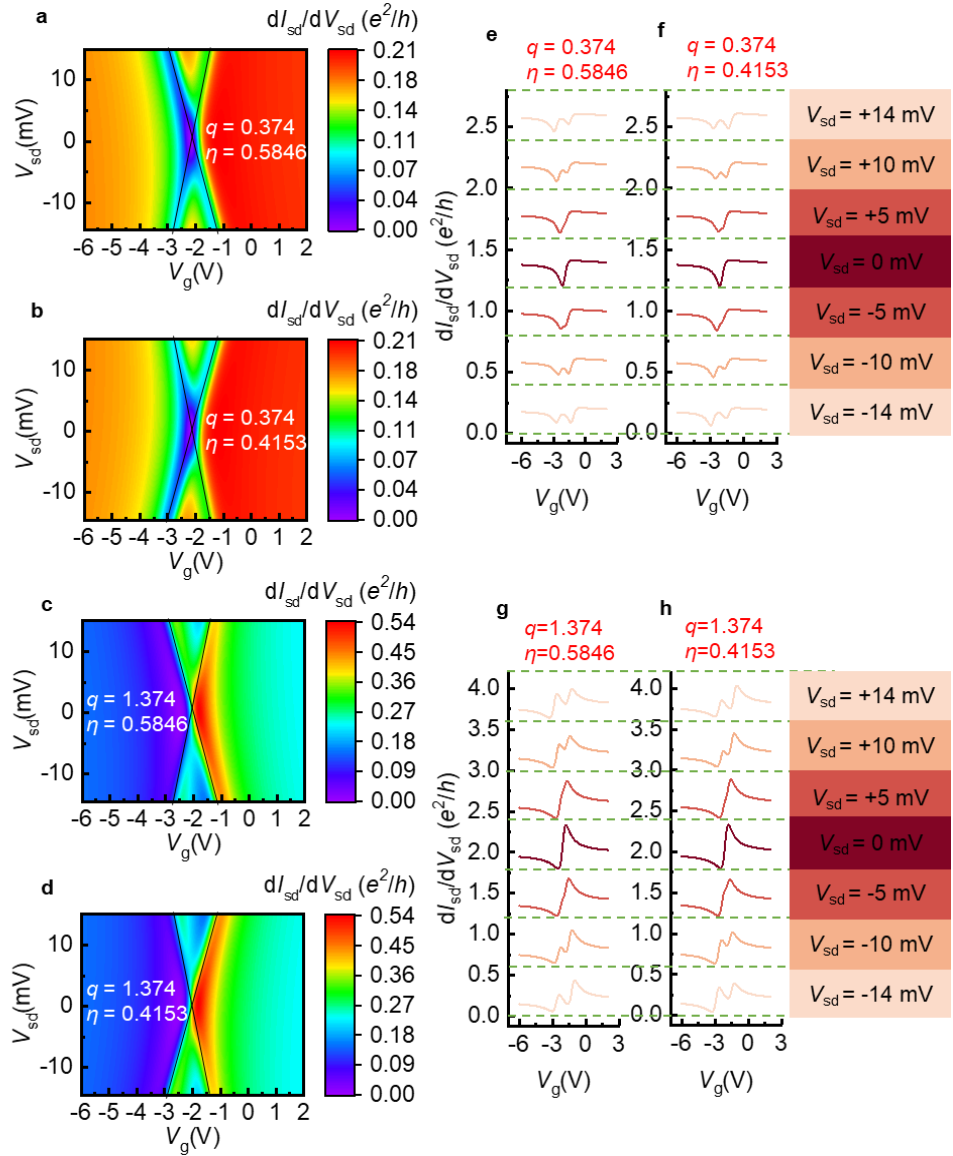

**Ouyang et al. Figure S3**

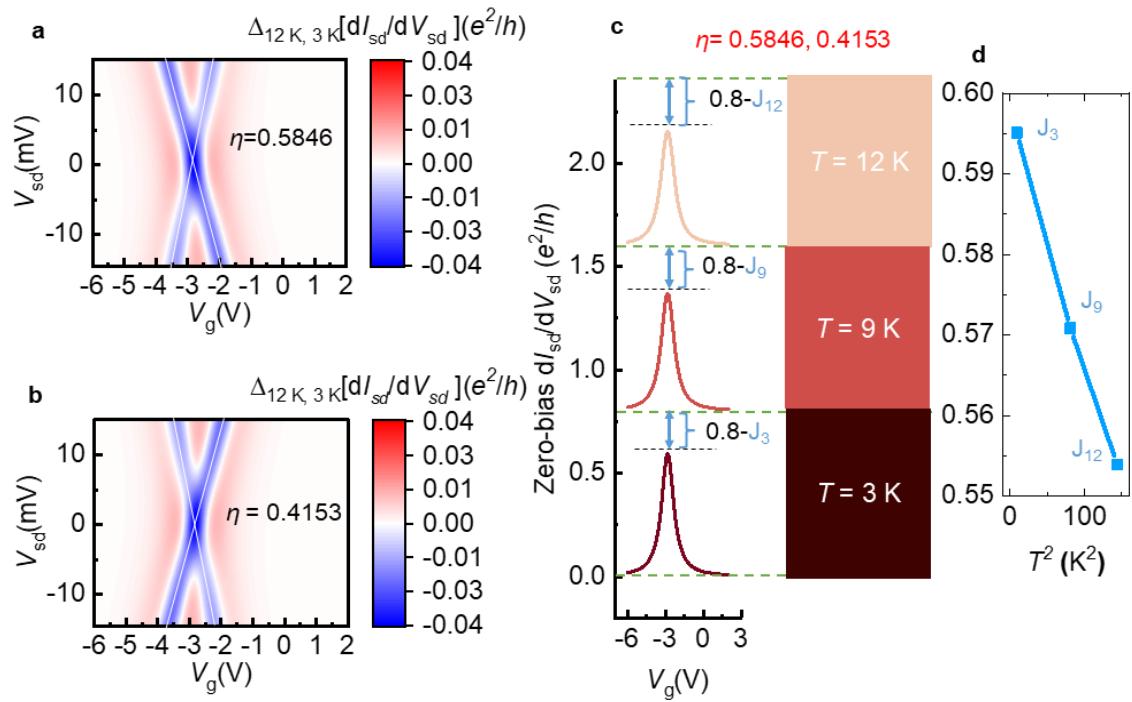

Ouyang et al. Figure S4

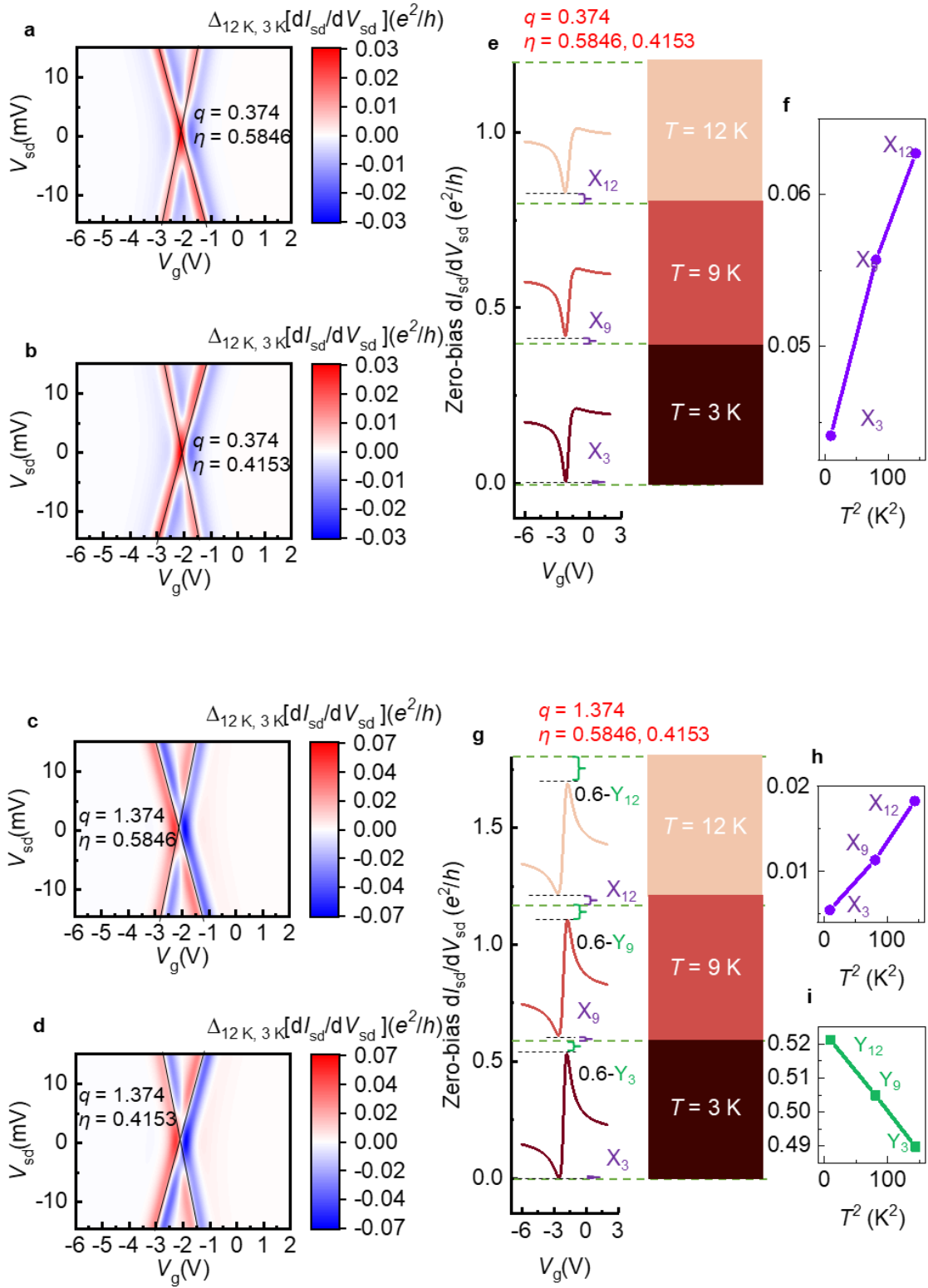

**Ouyang et al. Figure S5**

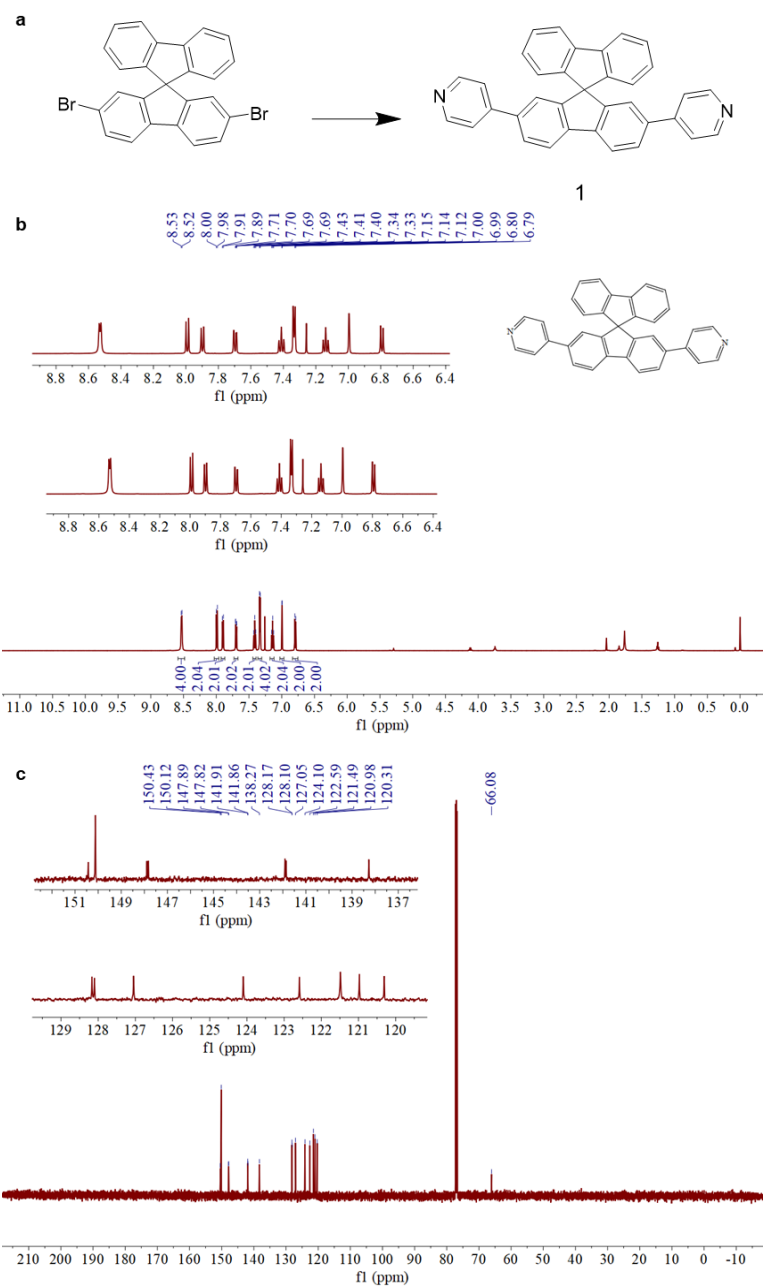

Ouyang et al. Figure S6

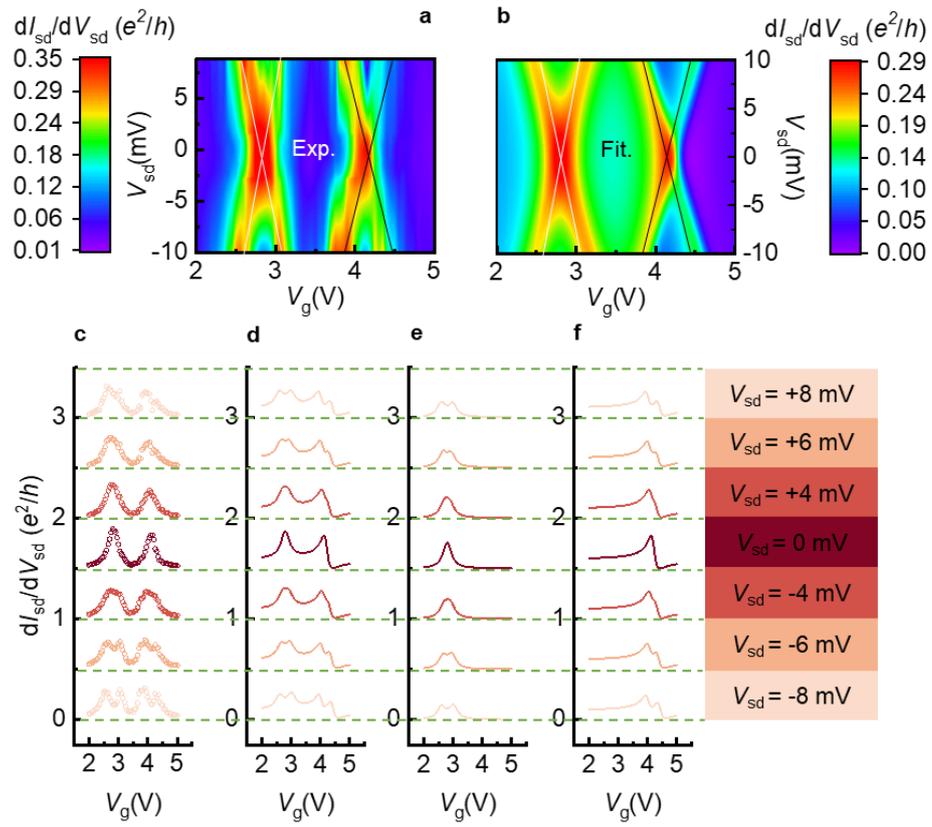

**Ouyang et al. Figure S7**